\documentclass[%
superscriptaddress,
preprint,
showpacs,
 amsmath,amssymb,
 aps,
pre
]{revtex4-1}

\usepackage{graphicx}
\usepackage{dcolumn}
\usepackage{bm}
\usepackage{hyperref}

\allowdisplaybreaks 

\begin{document}


\title{Bus bunching as a synchronisation phenomenon}

\author{Vee-Liem Saw}
\affiliation{Division of Physics and Applied Physics, School of Physical and Mathematical Sciences, 21 Nanyang Link, Nanyang Technological University, Singapore 637371}
\affiliation{Data Science and Artificial Intelligence Research Centre, Block N4 \#02a-32, Nanyang Avenue, Nanyang Technological University, Singapore 639798}
\author{Ning Ning Chung}
\affiliation{Complexity Institute, 61 Nanyang Drive, Nanyang Technological University, Singapore 637335}
\author{Wei Liang Quek}
\affiliation{Division of Physics and Applied Physics, School of Physical and Mathematical Sciences, 21 Nanyang Link, Nanyang Technological University, Singapore 637371}
\author{Yi En Ian Pang}
\affiliation{Division of Physics and Applied Physics, School of Physical and Mathematical Sciences, 21 Nanyang Link, Nanyang Technological University, Singapore 637371}
\author{Lock Yue Chew}
\email{lockyue@ntu.edu.sg}
\affiliation{Division of Physics and Applied Physics, School of Physical and Mathematical Sciences, 21 Nanyang Link, Nanyang Technological University, Singapore 637371}
\affiliation{Data Science and Artificial Intelligence Research Centre, Block N4 \#02a-32, Nanyang Avenue, Nanyang Technological University, Singapore 639798}
\affiliation{Complexity Institute, 61 Nanyang Drive, Nanyang Technological University, Singapore 637335}
%

\date{\today}

\begin{abstract}
Bus bunching is a perennial phenomenon that not only diminishes the efficiency of a bus system, but also prevents transit authorities from keeping buses on schedule. We present a physical theory of buses serving a loop of bus stops as a ring of coupled self-oscillators, analogous to the Kuramoto model. Sustained bunching is a repercussion of the process of phase synchronisation whereby the phases of the oscillators are locked to each other. This emerges when demand exceeds a critical threshold. Buses also bunch at low demand, albeit temporarily, due to frequency detuning arising from different human drivers' distinct natural speeds. We calculate the critical transition when \emph{complete phase locking} (full synchronisation) occurs for the bus system, and posit the critical transition to \emph{completely no phase locking} (zero synchronisation). The intermediate regime is the phase where clusters of partially phase locked buses exist. Intriguingly, these theoretical results are in close correspondence to real buses in a university's shuttle bus system.
\end{abstract}

\maketitle


\section{Introduction}

A self-oscillator is a unit with an internal source of energy (to overcome dissipation) that continuously and autonomously performs rhythmic motion. It is stable against perturbations on its amplitude but neutrally stable over perturbations on its phase. The latter allows an array of coupled self-oscillators to undergo phase synchronisation, as individual members affect others' phases through their nonlinear interactions \cite{Syn03}. This simple framework has provided a paradigm of immense utility for investigating various synchronisation phenomena, comprising circadian rhythms \cite{Winfree67,circa76}, neurons \cite{Neuron07}, Josephson junctions \cite{Joseph98,Joseph03,Joseph07}, a raft of other physical, chemical, social systems \cite{RMP05}, as well as complex networks \cite{ComplexNetworks16}. Plenty of research has been devoted to this area with rigorous mathematical treatments \cite{Syn03}. In particular, coupled self-oscillators exhibit phase transition: \emph{synchronisation emerges if coupling exceeds a critical threshold} \cite{Syn03,Kura75,Kura84,Stro91,Stro00}.

Dynamics of bus bunching, in contrast, is not as extensively studied by a physics approach. Buses arriving at bus stops in bunches forces commuters to face extended waiting times; and whilst indubitably undesirable, such occurrences are stable \cite{Newell64,Chapman78,Powell83,Bin06,Gers09,Daganzo09,Bell10,Cats11,Gers11,Bart12,WideDoor,Steward14,Sun14,Tirachini14,Gers15,Moreira16,Geneidy17,Wang18}. Bus bunching is a physical phenomenon of a complex socio-technological system. Hitherto, it is distinct from the concept of oscillator synchrony. Nevertheless, are there correspondences between them? Here, this paper establishes these connections and reveals the entrainment mechanisms of buses serving a loop of bus stops which underlie major bus routes at the heart of cities across the globe. Although many of the previous work on this subject had identified this problem and the inevitable occurrence of bunching \cite{Newell64,Chapman78, Powell83,Gers09,Bell10}, we have not encountered an approach based on a physical mechanism for synchronisation of coupled oscillators. Instead, some research dealt with potential strategies towards nullifying bus bunching by holding or delaying some buses, with some extending to adaptive and dynamic controls according to real-time situations \cite{Bin06,Daganzo09,Cats11,Gers11,Bart12,Moreira16,Wang18}, as well as exploring the effects of buses with wide doors \cite{WideDoor,Steward14,Geneidy17} or engineering the locations of bus stops along the bus routes \cite{Tirachini14}.

We present our physical theory of a bus loop system inspired by the Kuramoto model of coupled oscillators \cite{Kura75,Kura84} in the next section, followed by extensive simulation results in Section 3. Section 4 is devoted to real data from a university's bus loop system, which turn out to agree with the predictions of the theory. Some discussion concludes the paper, and technical derivations are given as supplementary information.

\section{Analytical theory}

\begin{figure}
\centering
\includegraphics[width=13cm]{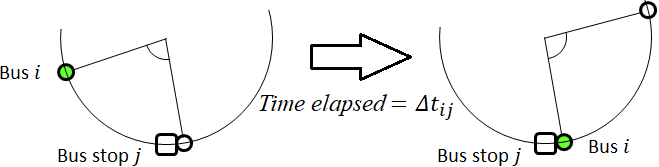}
\caption{Time headway $\Delta t_{ij}$ (temporal phase difference) with respect to bus stop $j$ between bus $i$ and the bus immediately ahead: the time interval between the moment the latter leaves bus stop $j$ and the moment bus $i$ leaves bus stop $j$.}
\label{fig1}
\end{figure}

Consider a bus system comprising $N$ buses serving $M$ bus stops in a loop. Each bus $i$, where $i\in\{1,\cdots,N\}$, is a self-oscillator: It has its own engine, fuel, and autonomy to travel along the loop; being human-driven, or driverless in the near future. Motion of bus $i$ is independent of its position or phase $\theta_i\in[0,2\pi)$ on the loop; it always moves with its natural (angular) frequency $\omega_i=2\pi f_i=2\pi/T_i$. If traffic slows it or if it momentarily accelerates, after that it just continues with $\omega_i$ without correcting for that phase perturbation (neutral stability). Without bus stops, bus $i$ loops around with $\omega_i$, oblivious and unaffected by other buses which can overtake. With bus stop $j$ present, where $j\in\{1,\cdots,M\}$, each bus $i$ must spend a stoppage $\tau_{ij}$ to board/alight passengers. Here, \emph{stoppage} refers to a bus being interrupted from its motion, as it has to stop at the bus stop to allow passengers to board or alight. We consider loading as the dominant process during $\tau_{ij}$, with passengers alighting simultaneously via different doors. So, this stoppage $\tau_{ij}$ is due to $P_j:=$ number of people at bus stop $j$; and $l:=$ loading rate onto the bus, i.e.\ $\tau_{ij}=P_j/l$. The loading rate $l$ is the same for each bus $i$ at each bus stop $j$. Bunched buses share the load equally, and would leave the bus stop simultaneously when everybody has been picked up. Furthermore, $P_j$ depends on the time headway $\Delta t_{ij}$ between bus $i$ and the bus immediately ahead (temporal phase difference), together with the average rate of people arriving at bus stop $j$, denoted by $s_j$. This time headway $\Delta t_{ij}$ (Fig.\ \ref{fig1}) is defined as the time interval between the moment the bus ahead leaves bus stop $j$ (so  $P_j$ resets to $0$) and the subsequent moment when bus $i$ leaves bus stop $j$ (resetting $P_j$ to $0$). Hence $P_j=s_j\Delta t_{ij}$, and overall:
\begin{align}\label{tau}
\tau_{ij}=k_j\Delta t_{ij}
\end{align}
gives the stoppage of bus $i$ at bus stop $j$ as a function of $\Delta t_{ij}$. The quantities $k_j:=s_j/l<1$ are parameters that determine the strength of the coupling amongst buses due to bus stops, as they determine how long a bus has to stop at the bus stop, together with $\Delta t_{ij}$. In other words, the strength of $k_j$ would magnify the effect of $\Delta t_{ij}$ in slowing down the buses. Later in Eq.\ (\ref{averageomega}), we see that $k_j$ appears in the expression for the average angular velocity for a bus where a higher $k_j$ would cause it to experience lower average angular velocity, and vice versa.

We summarise the dynamics of our model for the bus loop system as follows, where we would also implement in our simulations in Section 3:
\begin{enumerate}
\item A bus moves at its natural speed if it is not at a bus stop.
\item A bus would stop to allow alighting and boarding (simultaneously) if it is at a bus stop.
\item Suppose a bus is currently at a bus stop. If there is nobody on the bus who wishes to alight and there is nobody at that bus stop waiting to board, then the bus leaves and moves with its natural speed.
\item People are continually arriving at bus stop $j$ (one person at a time), at a rate of $s_j$ people per second. Equivalently, one person arrives at bus stop $j$ every $1/s_j$ seconds. (In our simulations in Section 3, we use the latter.)
\item People board or alight a bus at a rate of $l$ people per second.
\item For every person who boards a bus, we assign its destination to be the bus stop which is antipodally located (or near antipodal, if there is an odd number of buses).
\item Bunched buses share the load equally.
\end{enumerate}

This model describes \emph{a ring of discrete-local-unidirectionally coupled self-oscillators} for this bus loop system. The coupling is discrete as it only happens at bus stops; local because $\tau_{ij}$ only depends on the time headway from the bus immediately ahead; unidirectional due to only the leading bus directly affecting the trailing bus. Nevertheless, during the process where one bus bunches into another, these two buses directly affect each other i.e. bunched buses are bidirectionally coupled, since these buses have their load shared. A group of bunched buses can be regarded as \emph{one independent unit}, so  overall the whole system is viewed as having each independent unit unidirectionally affecting the independent unit directly behind it. In our simulations below in Section 3 based on the model described above, each bus moves autonomously as a self-oscillator. Evidently, the unidirectional effect is manifested by the fact that the number of people at the bus stop depends on how long ago the bus immediately ahead has left the bus stop. On the other hand, bunched buses mutually affect each other via their sharing of the load. This is also how real bus systems behave, for example in Section 4 below on a university campus loop shuttle bus service.

Note that: $1)$ $\Delta t_{ij}$ and $\tau_{ij}$ are \emph{directly measurable physical quantities}; $2)$ $\tau_{ij}$ depends on $\Delta t_{ij}$ and $k_j$, but not explicitly on $\omega_i$. Point (2) implies that inevitably present stochasticity of $\omega_i$ in real systems \emph{do not affect the relationship between $\Delta t_{ij}$ and $\tau_{ij}$}: If $\omega_i$ is slower, then $\Delta t_{ij}$ and correspondingly $\tau_{ij}$ [via Eq.\ (\ref{tau})] are increased. In addition, multiple (arbitrarily located) bus stops between bus $i$ and the bus ahead would certainly delay it but \emph{not affect the relationship between $\Delta t_{ij}$ and $\tau_{ij}$}, i.e.\ this analytical treatment applies to any $N, M = 0, 1, 2, \cdots$. Hence, we can employ Eq.\ (\ref{tau}) to empirically determine $k_j$ by measuring $\tau_{ij}$ and $\Delta t_{ij}$. Intriguingly, whilst Eq.\ (\ref{tau}) is linear, the coupling dynamics amongst buses is nonlinear. This is manifested by the average (angular) velocity of bus $i$ over a time interval $\eta+\tau_{ij}$ (where $\eta$ excludes stoppage) as it traverses bus stop $j$:
\begin{align}\label{averageomega}
\left\langle\frac{d\theta_i}{dt}\right\rangle = \omega_i\left(1-\frac{1}{1+\eta/k_j\Delta t_{ij}}\right),
\end{align}
i.e.\ $\langle d\theta_i/dt\rangle$ has a coupling term $-\omega_i/(1+\eta/k_j\Delta t_{ij})$ [a function of phase difference]. This is analogous to the Kuramoto model for synchronisation of an array of coupled self-oscillators whereby Eq.\ (\ref{averageomega}) has coupling term of the form $K\sum{\sin{\Delta\theta_{ij}}}$ \cite{Kura75,Kura84}. Unlike the \emph{original} Kuramoto model where coupling is a mean field globally contributed by every self-oscillator and is continuous, buses experience discrete coupling at bus stops (pulses), which is local (depending only on the bus immediately ahead) and unidirectional (instead of mutual).

Incidentally, there are various extensions to myriads of systems involving local coupling, different lattice arrangements of the oscillators, selective directional coupling, weighted and time-varying coupling, etc. carrying the name ``Kuramoto'' \cite{Neuron07,Joseph98,Joseph03,Joseph07}. These mainly still have a sine term in the coupling, highlighting the $2\pi$-periodicity with respect to phase difference. The Kuramoto model, however, assumes \emph{weak} coupling with several other approximations/simplifications (see for example, Ref. \cite{Syn03}). There are in fact, more general coupling functions \cite{Daido94,Craw95}. Our derivation for the bus system here is \emph{exact}.

Buses bunch in two ways: 1) frequency detuning; 2) phase locking. Due to frequency detuning, a fast bus catches a slow one, overtakes, then escapes: The system exhibits \emph{periodic bunching}. This is always present due to human drivers' distinct $\omega_i$. In contrast, strong coupling during high demand causes phase synchronisation where some adjacent pair of buses' \emph{spatial phase difference}, $\Delta\theta$, becomes small and \emph{bounded}. Hence, we classify the bus system's dynamics into two phases (note the dual usage of ``phase''): (a) \emph{lull}, where periodic bunching occurs due to frequency detuning; and (b) \emph{busy}, where phase locking (sustained bunching) forms at high demand. Frequency detuning is a double-edge sword: It is a source of non-synchrony in an ensemble of oscillators (purportedly preferable in undoing clustering), but simultaneously prevents stable constant $\Delta\theta$ (which is inappropriate).

Coupled self-oscillators generally experience phase synchronisation, given sufficiently strong coupling. For instance, the Kuramoto model provides an exact analytical treatment for \emph{infinitely} many oscillators with natural frequencies given by a unimodal symmetric distribution $g(\omega)$. By considering the density distribution of these oscillators over the loop, the critical transition for synchronisation occurs at $K_c=2/\pi g(0)$ \cite{Kura75,Kura84,Stro91,Stro00}. For finite $N=2,3,\cdots$ buses, we derive an analytic expression for the critical $k_c$ where phase transition occurs between \emph{complete} (CPL) and \emph{partial phase locking} (PPL). Suppose each of the $M$ bus stops has equal people arrival rate $s$ (so $k:=s/l$) and they are perfectly staggered. This $k_c$ is:
\begin{align}\label{criticalk}
k_c(N)=\frac{1}{M}\sum_{i=1}^{N-1}\left(1-\frac{\omega_N}{\omega_i}\right),
\end{align}
where the $N$ buses have natural frequencies $\omega_1>\cdots>\omega_N$, respectively [derivation of Eq.\ (\ref{criticalk}) in the supplementary information]. Buses in CPL would bunch as a single unit at each bus stop. After picking up everybody, they leave simultaneously with the faster ones pulling away, but get bunched completely at the next bus stop due to high demand. They are thus in equilibrium, with this sequence of events repeating at every bus stop. Eq.\ (\ref{criticalk}) gives a dimensionless quantity, depending on dimensionless ratios $\omega_N/\omega_i=f_N/f_i=T_i/T_N$.

An $N$-bus system is CPL if $k>k_c(N)$. There is another critical $\overline{k}\leq k_c(N)$ marking the phase transition between PPL and \emph{no phase locking} (NPL). If $k<\overline{k}$, then all buses do not experience phase locking but occasionally bunch due to frequency detuning. PPL is the regime $\overline{k}<k<k_c(N)$. The lull phase refers to NPL where bunching only occurs due to frequency detuning; the busy phase includes both PPL and CPL where at least one sustained bunching is present. The case $N=2$ is special: $\overline{k}=k_c(2)$ with only NPL (lull) and CPL (busy). In real bus systems, we should never have to encounter CPL where all $N$ buses bunch together, as this would be a highly inefficient system.

\section{Simulations}

\begin{figure}
\centering
\includegraphics[width=16cm]{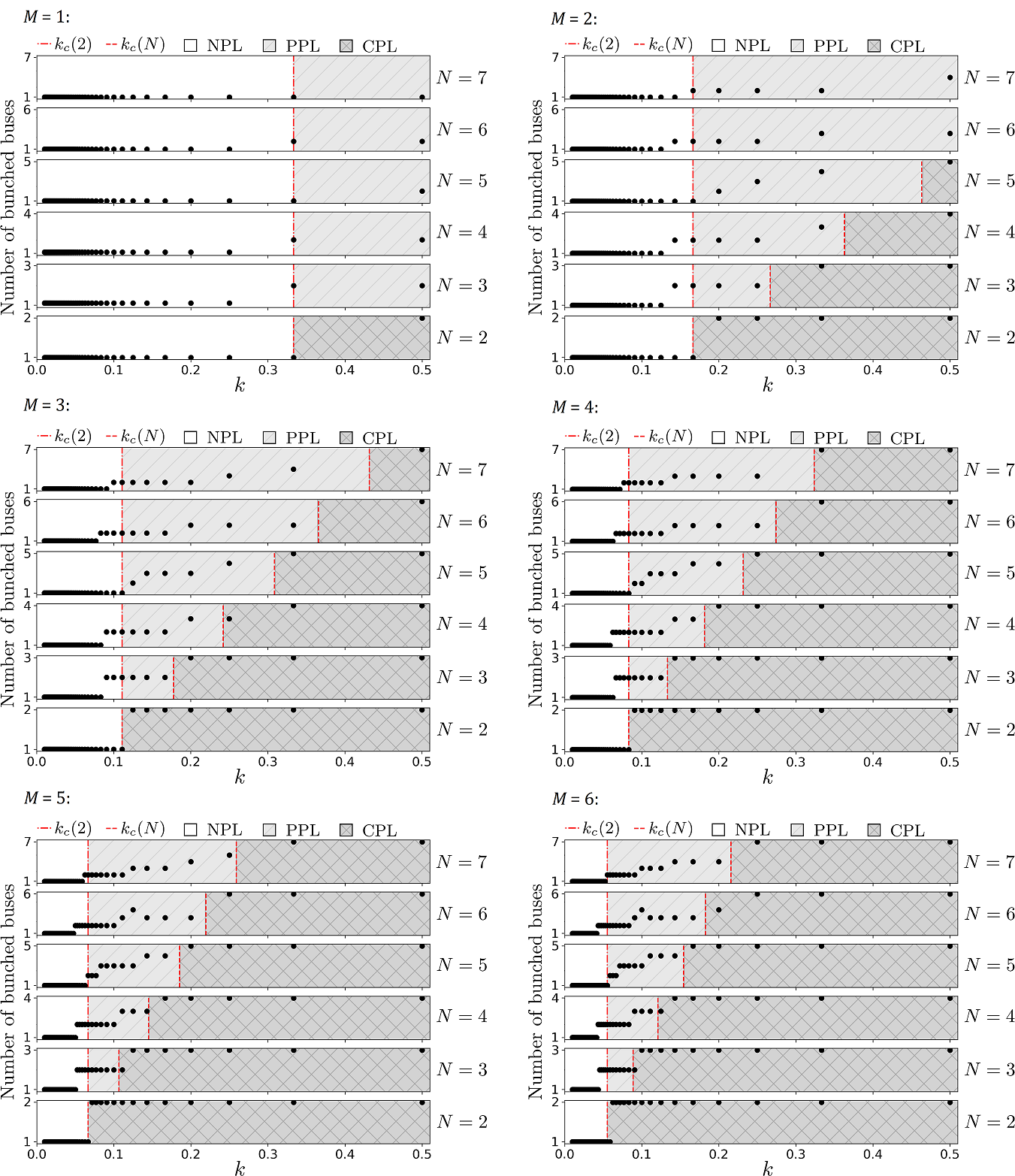}
\caption{Simulation results of $N=2,3,4,5,6,7$ buses, respectively, serving $M=1,2,3,4,5,6$ bus stops in a loop, respectively: The number of bunched buses increases stepwise as $k$ increases. CPL is the region $k>k_c(N)$, where \emph{all} $\Delta\theta_\textrm{max}\sim0^\circ$. NPL is observed to occur when $k<\overline{k}\sim k_c(2)$, where \emph{all} $\Delta\theta_\textrm{max}\sim180^\circ$. PPL for $N>2$ lies in between these extremes, as demarcated by the two vertical lines in the graphs. These vertical lines are analytical values from Eq.\ (\ref{criticalk}).}
\label{fig2}
\end{figure}

\begin{figure}
\centering
\includegraphics[width=16cm]{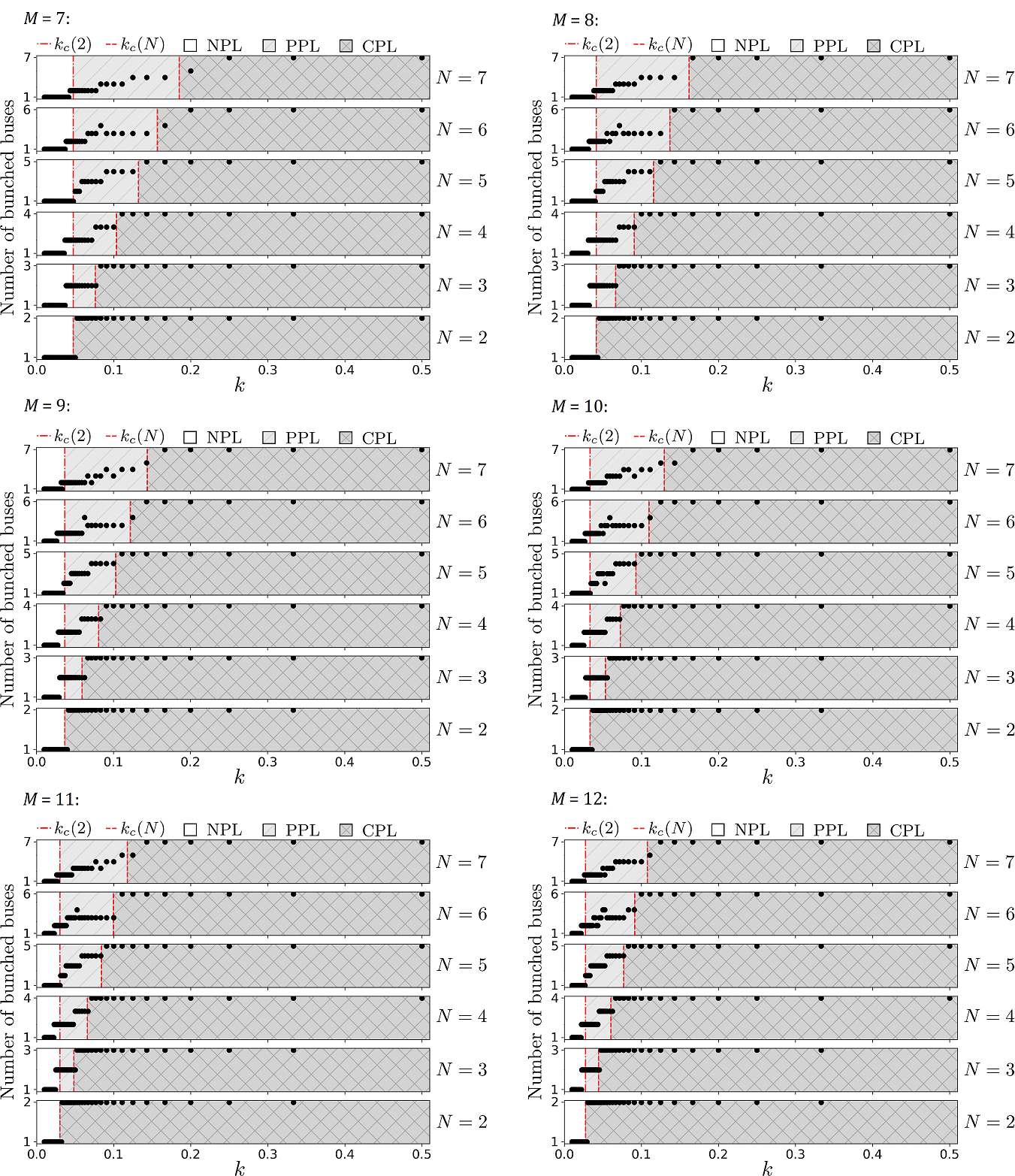}
\caption{Fig.\ \ref{fig2} continued, with $M=7,8,9,10,11,12$ bus stops, respectively.}
\label{fig3}
\end{figure}

We carry out extensive simulations to determine the various degrees of local clustering of sustained bunching. This is done for $N=2,3,4,5,6,7$ buses, respectively, with natural frequencies given in Table \ref{table1}, serving $M=1,2,\cdots,12$ staggered bus stops, respectively, in a loop. From Eq.\ (\ref{criticalk}), theoretical values of $k_c(N)$ for the transition between PPL and CPL are calculated and compared directly with the simulation results in Figs.\ \ref{fig2}-\ref{fig3}. Note that Table\ \ref{table1} only shows the analytical values of $k_c(N)$ for $12$ bus stops, as $M=12$ is of particular interest with regards to a real bus loop system (see the next section).

For each bus, its $\Delta\theta$ is the phase difference between itself and the bus immediately ahead of it. We run the simulations over a very long time and measure the maximum value of $\Delta\theta$ in the steady state part of the simulation. This maximum value of $\Delta\theta$ is $\Delta\theta_\textrm{max}$. If there are $N$ buses, then there are $N$ such $\Delta\theta_\textrm{max}$. Generally, an $N$-bus system has $N-1$ independent local phase differences. If there is local clustering between a pair of buses, then its $\Delta\theta_\textrm{max}$, is small ($\sim0^\circ$); whereas large $\Delta\theta_\textrm{max}$ ($\sim180^\circ$) signifies no local phase locking since it is able to open up a large phase difference. The number of small $\Delta\theta_\textrm{max}$ represents the degree of local clustering. As shown in Figs.\ \ref{fig2}-\ref{fig3}, increasing $k$ would increase the degree of local clustering \emph{stepwise} until $k_c(N)$ where CPL emerges with all $\Delta\theta_\textrm{max}\sim0^\circ$. Our simulations register slightly higher $k_c(N)$ for the critical transition between PPL and CPL than the analytical results given by Eq.\ (\ref{criticalk}). This is because people are \emph{discrete}. If ``0.9 person'' is due to arrive, then the bus(es) would leave instead of loading that ``0.9 person''. So, slightly stronger coupling is required to keep them bunched.

Whilst we do not have the analytical calculation for the critical $\overline{k}$ for $N>2$ that marks the transition between NPL and PPL (because these are non-equilibria with unbunched buses continually affecting the bunched pair), simulations suggest that this phase transition occurs at $\overline{k}\sim k_c(2)$, for that respective $M$. Thus, $k_c(2)<k<k_c(N)$ is a reasonable indication of PPL. In fact, as $k_c(2)$ for $N=2$ represents the critical value where the two buses would exhibit persistent bunching, for $N>2$ this same value $k_c(2)$ gives an order of magnitude when two of these $N$ buses may begin to experience persistent bunching as well, hence $\bar{k}\sim k_c(2)$. In the real world however, other perturbations like non-constant speed, traffic influences, as well as stopping at junctions may play a part in affecting the dynamics. Note that although $k<\overline{k}$ such that all $\Delta\theta_\textrm{max}=180^\circ$, simulations reveal transient local clustering periodically appearing.


\begin{table}
\centering
\begin{tabular}{|c|c|c|}
\hline
$N$ & $f_i$ (mHz) & $k_c$\\
\hline
$2$ & $1.39, 0.93$ & $0.028$\\
$3$ & $1.39, 1.16, 0.93$ & $0.045$\\
$4$ & $1.39, 1.24, 1.08, 0.93$ & $0.061$\\
$5$ & $1.39, 1.24, 1.16, 1.08, 0.93$ & $0.077$\\
$6$ & $1.39, 1.31, 1.24, 1.08, 1.00, 0.93$ & $0.091$\\
$7$ & $1.39, 1.31, 1.24, 1.16, 1.08, 1.00, 0.93$ & $0.108$\\
\hline
\end{tabular}
\caption{Table of $k_c$ for various $N$ buses with different $f_i\in[0.93,1.39]$ mHz. The analytical values of $k_c$ from Eq.\ (\ref{criticalk}) are for the case with $M=12$ bus stops in a loop.}\label{table1}
\end{table}

\begin{figure}
\centering
\includegraphics[width=16cm]{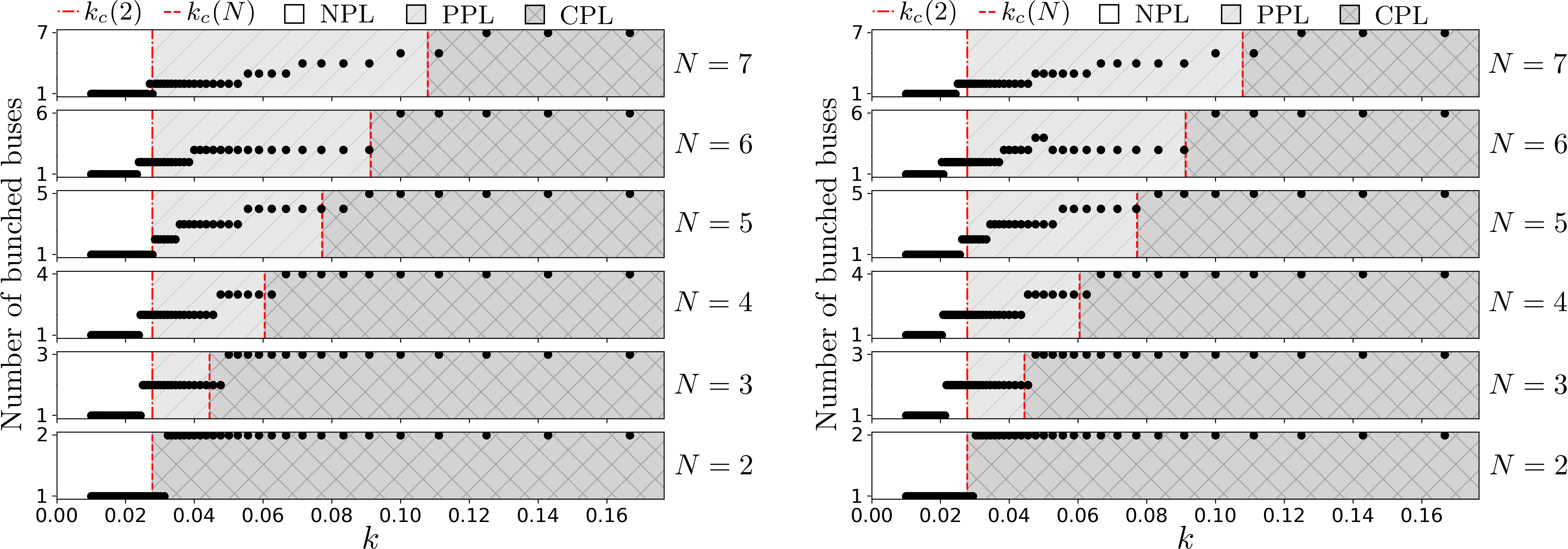}
\caption{$M=12$ bus stops, with the entire set of frequencies being doubled (left) and the entire set of frequencies being halved (right).}
\label{fig4}
\end{figure}

The simulations with different number of bus stops $M=1,2,\cdots,12$ affirms that $M$ only acts as a multiplier of the coupling strength, whilst the qualitative features are essentially identical. Furthermore, we also run separate simulations with $M=12$ bus stops where the set of frequencies in Table\ \ref{table1} are all doubled, as well as all halved, presented in Fig. \ref{fig4}. Once again, the qualitative features are essentially identical to Figs.\ \ref{fig2}-\ref{fig3}.

\section{Data analytics on the Nanyang Technological University campus buses}

\begin{figure}
\centering
\includegraphics[width=16cm]{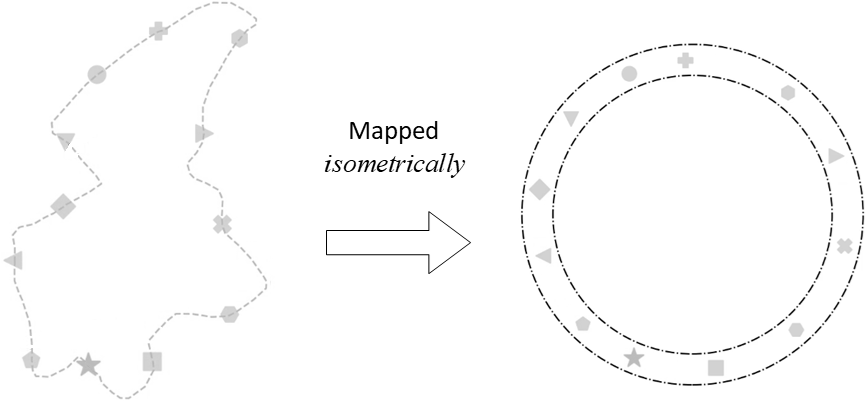}
\caption{Left: A map of the \emph{Blue route} of the Nanyang Technological University, Singapore campus shuttle buses. Buses go clockwise. Right: An isometric map (which preserves distance) of the actual route to a circle, showing that the $M=12$ bus stops are reasonably staggered around the loop.}
\label{fig5}
\end{figure}

We can apply our physical theory of buses as a synchronisation phenomenon to our Nanyang Technological University (NTU) Campus Buses \cite{NTUnews}. We collect and process positional data for buses on the \emph{Blue} route. Live data can be found here: \url{https://baseride.com/maps/public/ntu/}. There are $M=12$ bus stops along a loop within our campus (see Fig.\ \ref{fig5}), usually served by $3-4$ buses ($1-2$ during off-peak hours), with $7-8$ buses over busy periods (weekdays $8.30-10.30$am). These human-driven buses do not have identical natural frequency, but take an average of $12-18$ minutes to complete a loop without stoppages, i.e.\ $f_i\in[0.93,1.39]$ mHz. This frequency range forms the basis for the prescribed $f_i$ in Table \ref{table1}. Bus bunching is a highly observed phenomenon, with $4-5$ buses bunching together being a ubiquitous sight in NTU. This perennial and notorious issue on bus bunching affects students, staff and faculty members who live, work and play in NTU.

\begin{figure}
\centering
\includegraphics[width=16cm]{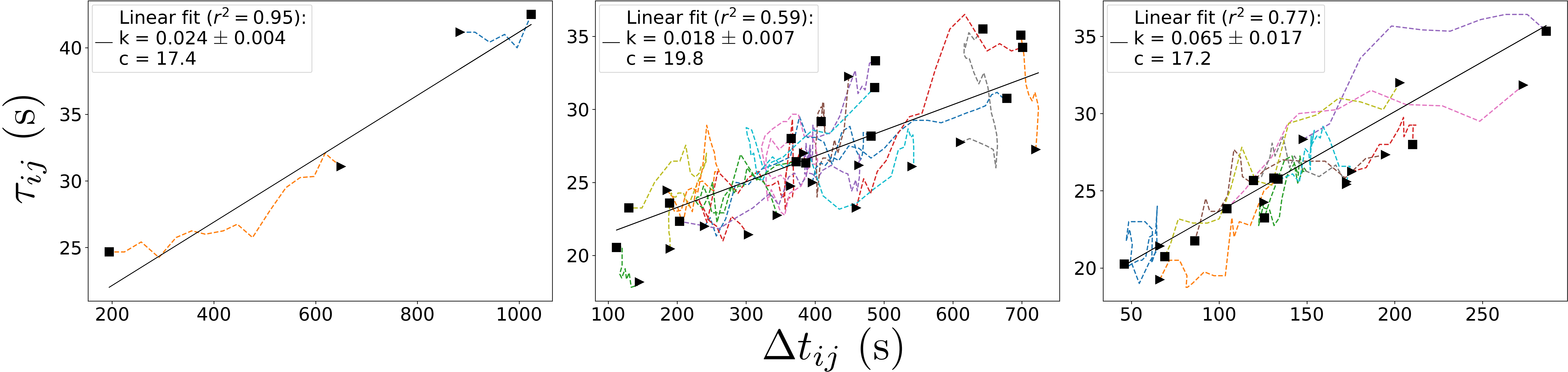}
\caption{Graphs of $\tau_{ij}$ versus $\Delta t_{ij}$ (loop averages) of our NTU Campus Buses, for the three situations described in the text. Left and middle are in the \emph{lull phase}; right is in the \emph{busy phase}. A time series curve tracks one bus throughout the measured period, plotting its loop average values (over all $12$ bus stops) of $\Delta t_{ij}$ and $\tau_{ij}$ as it moves from one bus stop to the next. Each time series starts from $\blacktriangleright$ and ends at $\blacksquare$.  The number of time series curves corresponds to the number of buses being tracked. There are $2$ buses tracked in the left graph, $5\times3=15$ buses tracked in the middle graph, and $10$ buses tracked in the right graph.}
\label{fig6}
\end{figure}

Since bus stops have heterogeneous people arrival rates, a fair global measure is obtained by averaging quantities over all $12$ bus stops, i.e.\ to obtain \emph{loop averages} as the system evolves. As time goes on, a bus would move from one bus stop to the next bus stop. (Each bus stop is a location where people board/alight the bus.) At the current bus stop, we take the average over the $12$ bus stops in the loop, i.e. from the past $11$ bus stops up to that present bus stop. When the bus moves to the next bus stop, we correspondingly shift the average over all $12$ bus stops from the past one revolution to that new bus stop, and so on. Therefore, we calculate loop averages of $\Delta t_{ij}$ and $\tau_{ij}$ to plot the time series for each bus $i$, thus deducing the average $k$ via Eq.\ (\ref{tau}). In doing so, we track these averages for each bus with time as the system evolves, as the bus moves from one bus stop to the next bus stop. This tracking over time as the system evolves would form curves connecting the discrete points $(\Delta t_{ij}, \tau_{ij})$ in a plot of $\tau_{ij}$ versus $\Delta t_{ij}$ as shown in Fig.\ \ref{fig6}, one curve for each bus being tracked.

Fig.\ \ref{fig6} shows the time series of our NTU buses recorded over the entire working week $16-20^\textrm{th}$ of April, $2018$, clearly depicting the two phases \emph{lull} and \emph{busy}, as described in Section 2. The first (left panel) is a quintessential example of two initially antipodally-spaced buses, ending up bunching due to frequency detuning. Their average natural frequencies are $0.94$ mHz and $1.14$ mHz, respectively. These average natural frequencies fluctuate, which effectively contribute towards greater frequency detuning. This event occurred at $9.32-10.33$pm on the $16^\textrm{th}$. In fact, the two curves corresponding to tracking each of the two buses vividly indicate that for one of them, its time headway $\Delta t_{ij}$ and stoppage $\tau_{ij}$ progressively get \emph{smaller} as time goes on, whilst those for the other bus progressively get \emph{larger} --- which signify imminent bunching between these two buses.

Next (middle panel) is a collection of time series between $4-5$pm during those $5$ days, where $3$ buses were serving during this hour ($15$ time series in total). Before $4$pm and after $5$pm, buses are rested/replaced, i.e. the system is tweaked or non-isolated. This $4-5$pm is an interval where the system is \emph{isolated}. This and the $2$-bus system in the left panel represent real-world examples of the lull phase where bunching occurs periodically with the fast bus pulling off after overtaking. The third (right panel) is a selection of 10 time series throughout those $5$ days from $9-10$am, served by $6-7$ buses during the morning rush hour. This is a typical situation of the busy phase where clusters of phase-locked buses are recorded. Here, we make a selection because these $6-7$ buses often overtake one another constituting a transient state which destroys the loop average sequence. Bunched buses share the load, so the loading rate $l$ gets multiplied --- which is not $k$ for a single bus. In addition, these are $10$ time series with highest demand, to determine a representative value for \emph{peak demand} throughout the week. Demand varies during morning rush hour because lectures in NTU start at the half-hour mark, so students travelling from campus residences to faculty buildings are likely to leave at preferred times. For the first two graphs (lull phase), overtaking is not as frequent and no time series is excluded.

The plots in Fig.\ \ref{fig6} signify a \emph{linear relationship} between $\tau_{ij}$ and $\Delta t_{ij}$ [Eq.\ (\ref{tau})], with expected real-world stochasticity. A source of deviation from Eq.\ (\ref{tau}) arises from disembarkation: Buses carrying many passengers may stop longer, although the time headway may be relatively small with few boarding. Also, if bus capacity is reached, excess passengers are ignored. Anyway, we can fit a straight line to obtain the gradient $k$ and the $\tau_{ij}$-intercept. The $\tau_{ij}$-intercept is not quite zero, but of the order of $+10$ seconds for all $\tau_{ij}$-$\Delta t_{ij}$ graphs. Note that buses have to wait for clear traffic before rejoining the road. This could only lead to an increase in measured $\tau_{ij}$ and never a decrease in $\tau_{ij}$. We wrote a computer programme to regularly probe the website \url{https://baseride.com/maps/public/ntu/} for positional data of the buses. However, the positional data on that website are only updated once in approximately every $10$ seconds, i.e these positional data would have a resolution of about $10$ seconds. Due to this rather coarse resolution, the increase due to waiting for clear traffic would be recorded as an increase by an order of $10$ seconds. More specifically, a bus might have waited for, say for example, either $1$ second, $2$ seconds, $5$ seconds, or $10$ seconds. These would all be registered as about $+10$ seconds, in addition to the actual time taken for boarding/alighting. If a bus waits for, say for example, either $11$ seconds, $12$ seconds, $15$ seconds, or $20$ seconds, then these would all be registered as about $+20$ seconds, in addition to the actual time taken for boarding/alighting. Buses would be waiting for clear traffic over a range of values, depending on the traffic conditions. So the $\tau_{ij}$-intercept in each graph in Fig.\ \ref{fig6} gives an average over $12$ bus stops, on how long a bus has to spend waiting for clear traffic before rejoining the road.

From the gradients of the fitted lines, $k=0.024\pm0.004$ for the first graph, $k=0.018\pm0.007$ for the second [breakdowns for the $5$ respective days ($4-5$pm) are: $0.021\pm0.007,0.028\pm0.005,0.015\pm0.006,0.011\pm0.005,0.024\pm0.004$], and $k=0.065\pm0.017$ for the third. \emph{Estimate}: Suppose $1$ person takes $1$ second to board/alight. If $10$ people approach a bus stop per minute at busy hours, compared to $1$ in $5$ minutes during lull times, these translate to $k\in[0.003,0.167]$. These estimates serve as extreme boundary values for $k$, and our results lie moderately within this range without being near either extreme ends.

The data analytics strongly suggest a classification of the bus system into distinct lull and busy phases, as mentioned earlier. Frequency detuning allows a fast bus to pull away after overtaking. Although it gets held up at subsequent bus stops, during lull periods its stoppage is short and its higher speed helps widen the spatial phase difference, $\Delta\theta$. However, when coupling gets beyond a critical value, the bus system transitions from lull to busy where clusters of phase-locked buses appear: whilst the fast bus escapes, it then experiences lengthy stoppage at the next bus stop and gets caught up. This sequence of events repeats at every bus stop. Since frequency detuning causes periodic bunching under weak coupling, they are also periodically staggered: $\Delta\theta_\textrm{max}=180^\circ$. If coupling gets beyond the critical threshold, phase locking emerges: $\Delta\theta_\textrm{max}\sim0^\circ$. Periodic bunching and phase locking are the two stable phases of human-driven bus systems, where their natural frequencies are different.

\begin{figure}
\centering
\includegraphics[width=13cm]{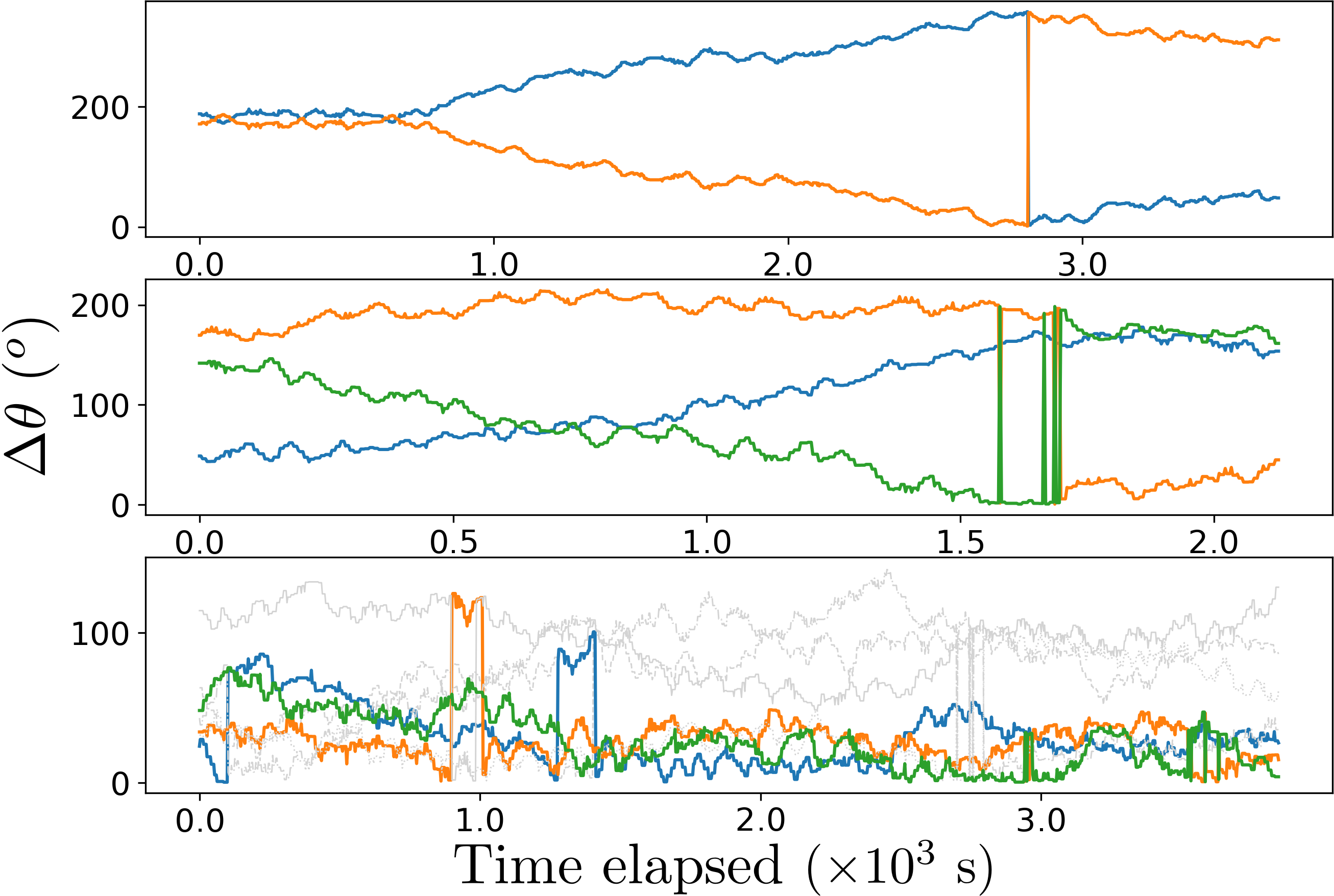}
\caption{Graphs of $\Delta\theta$ between adjacent NTU campus buses versus time elapsed. Each curve represents the spatial phase difference $\Delta\theta$ of a bus, as measured from the bus immediately ahead of it. Top: $2$ buses serving at $9.32-10.33$pm on the $16^\textrm{th}$; middle: $3$ buses serving at $4.32-5.07$pm on the $20^\textrm{th}$; bottom: $7$ buses serving at $8.56-10.13$am on the $19^\textrm{th}$ of April, 2018. The first two are in the \emph{lull phase} with frequency detuning causing a fast bus to chase, overtake, then pull away from a slow bus. The third is in the \emph{busy phase} where clusters of sustained phase-locked buses are visible. In the bottom graph, bunched buses are represented by thick coloured curves where $\Delta\theta$ is small, whilst other buses' $\Delta\theta$ are large (grayscale thin curves). Videos for these are given as supplementary information.}
\label{fig7}
\end{figure}

In summary, data from our NTU buses turn out to fit well according to our physical theory's predictions:
\begin{enumerate}
\item A $2$-bus scenario on Monday, $16^\textrm{th}$ of April, 2018 ($9.32-10.33$pm) is in the \emph{lull phase} where they are not phase locked, with measured $k=0.024\pm0.004$ being less than $k_c(2)=0.028$.
\item The cumulative $3$-bus scenarios for an entire working week, $16-20^\textrm{th}$ of April, $2018$ ($4-5$pm, before the evening rush hour) is also in the \emph{lull phase}, with measured $k=0.018\pm0.007$ being less than $k_c(3)=0.045$. This measured value is also less than $\overline{k}=k_c(2)=0.028$. No clustering is observed.
\item The cumulative $6$ to $7$-bus scenarios  for an entire working week, $16-20^\textrm{th}$ of April, $2018$ ($9-10$am, morning rush hour) is in the \emph{busy phase}, with measured $k=0.065\pm0.017$. Although less than $k_c(6)=0.091$ and $k_c(7)=0.108$ (thankfully, otherwise this would be a ridiculously inefficient system), it is much higher than $\overline{k}=k_c(2)=0.028$, indicating the presence of clusters of phase-locked buses.
\end{enumerate}

Fig.\ \ref{fig7} supports the theoretical deduction of these scenarios based on their measured $k$, by showing how $\Delta\theta$ between adjacent pairs of NTU buses evolve with time, with the loop isometrically (distance-preserving) mapped to a circle (Fig.\ \ref{fig5}). Recall that $\Delta\theta$ for each bus is measured with respect to the bus immediately ahead of it. In each graph, there are $N$ curves, one for each of the $N$ buses being tracked. These $\Delta\theta$ for each bus always add up to $360^\circ$, at every point in time. Note that there are jumps in the curves corresponding to overtakings amongst the various buses. When overtaking between two buses occurs, the corresponding two curves experience vertical jumps since the bus immediately ahead would change to a different bus. If a bus overtakes the bus immediately ahead of it, then before overtaking its $\Delta\theta$ goes to zero. After overtaking, its $\Delta\theta$ is measured with respect to a different bus that is now immediately ahead of it. This new bus immediately ahead of it could be anywhere around the loop, and so its $\Delta\theta$ would jump from $0^\circ$ to that value corresponding to measuring with respect to the new bus immediately ahead. Similarly for the bus that is being overtaken, its $\Delta\theta$ could be any value from $0^\circ$ to $360^\circ$ before being overtaken, depending on how far away the bus immediately ahead of it is. After it is overtaken, its $\Delta\theta$ drops to $0^\circ$ because the bus immediately ahead of it is the one that has just overtaken it.

\section{Discussion}

Let us now consider $N$ initially staggered buses with \emph{identical} $\omega=2\pi f=2\pi/T$ serving $M$ bus stops, as we anticipate a future when self-driving buses are programmed. It turns out that although this is an equilibrium where they can remain staggered, stability \emph{bifurcates} from neutrally stable (upon perturbations, the system does not return to the original configuration nor deviate away further) to unstable (upon perturbations, the system deviates away from the original state) at a critical $k=k_{c'}$:
\begin{align}\label{kcprime}
k_{c'}(N)=\frac{N\tau_\textrm{min}}{T}.
\end{align}
To derive Eq.\ (\ref{kcprime}), note that a bus must spend at least some minimum amount of time to board even one passenger. Let $\tau_\textrm{min}$ denote this minimum duration. If $l\tau_\textrm{min}$ is at least $P_\textrm{max}:=$ maximum number of people accumulated at each bus stop, then \emph{all buses would only spend the minimum stoppage $\tau_\textrm{min}$} to board them. When the $N$ buses are staggered, we have $\Delta t_{ij}\sim T/N$. Therefore, $P_\textrm{max}=sT/N$ and the critical transition is given by $l\tau_\textrm{min}=P_\textrm{max}=sT/N$, i.e. Eq.\ (\ref{kcprime}). For $k<k_{c'}$, the system is neutrally stable since all stoppages are $\tau_\textrm{min}$ even with small perturbations --- the buses do not end up bunching, and do not return to its original configuration. If all buses only spend the minimum time of $\tau_\textrm{min}$ at the bus stop, then a small perturbation on the phases of an initially staggered system of buses does not change the fact that they would all still spend the same $\tau_\textrm{min}$ at the bus stop. On the other hand, if $k>k_{c'}$, the staggered configuration is generally unstable and all buses end up bunching. Moreover, we have run simulations on several configurations of $N$ buses serving $M$ bus stops in a loop to test this and indeed observed in each case that the buses can remain staggered if $k<k_{c'}$, but would bunch if $k>k_{c'}$. These stability properties of an initially staggered configuration of buses are analogous to those of the original Kuramoto model \cite{Stro91}. Ref. \cite{Stro91} provided a rigorous mathematical treatment on the \emph{incoherent solution} to the Kuramoto system (i.e. where the infinitely many oscillators are randomly distributed around the loop), concluding that it is neutrally stable if the coupling strength is less than the critical value, but is unstable if the coupling strength exceeds the critical value. Intriguingly and unlike the original Kuramoto model, our simulations also show that the system can remain in staggered equilibrium for some values of $k>k_{c'}$ due to the \emph{wide-doors effect} allowing multiple passengers boarding simultaneously \cite{WideDoor,Steward14,Geneidy17}: If a bus can pick up $l$ people per time step, then picking up $1,2,\cdots,l$ people all require one time step, i.e. tiny perturbations are tolerable.

Let us compare the efficacy against bunching for bus systems with identical $\omega$ versus non-identical $\omega_i$. With $N=5$, $T=15$ minutes ($\omega=1.11$ mHz) and $\tau_\textrm{min}=5$ seconds (which includes deceleration and acceleration), then $k_{c'}(5)=0.028$. This is similar to $k_c(2)$ for $M=12$ bus stops with frequency detuning (Table\ \ref{table1}), but requires $5$ buses. With $N=2$, $k_{c'}(2)=0.011$ implies that a coupling strength with, say $k=0.020$, would cause bunching. Unlike with frequency detuning, bunched buses with identical $\omega$ stay bunched.

In conclusion, we can model complex real bus systems by a simple physical theory of self-oscillators coupled via Eq.\ (\ref{tau}), where the latter successfully captures essential underlying features of the former. Perhaps the most significant takeaway from this physical theory is, since \emph{self-oscillators can be entrained}, buses can be kept staggered by a system of periodic driving forces. As lucidly described in Ref. \cite{Syn03}, ordinary clocks these days need not be very accurate. A high-precision central clock can send periodic pulses (via radio signals) to entrain those clocks, safeguarding their accuracy. Similarly, we can design a set of central oscillators for the bus system that dictates the $N$ \emph{ideal phases} for the $N$ buses, whereby signals are sent out periodically to entrain them to remain staggered. In the real world, such central oscillators must continuously adapt to varying demands and traffic conditions. Hence, we are presently developing a smart central system using data science and artificial intelligence feeding on live demand, real-time traffic data, etc. \cite{Bin06,Moreira16}, thereby inhibiting bus bunching by driven entrainment.

\bibliography{Citation}

\begin{acknowledgments}
This work was supported by MOE AcRF Tier 1 (Grant No. RG93/15) and the Joint WASP/NTU Programme (Project No. M4082189).
\end{acknowledgments}

\section*{Author Contributions}
L.Y.C. designed research; V.-L.S., L.Y.C. and N.N.C. performed research; V.-L.S. collected and analysed data; V.-L.S., N.N.C. and Y.E.I.P. performed analytical study; V.-L.S. and W.L.Q. performed numerical simulations; V.-L.S. and L.Y.C wrote the manuscript; V.-L.S. prepared all figures.

\section*{Additional Information}
\subsection*{Supplementary information}
Supplementary information accompanies this paper. Videos can be found here:\\\url{https://www.youtube.com/playlist?list=PLZIj25fISvwOUj1ESCW0pkBbGMMKzk9Fc}
\subsection*{Competing Interests}
The authors declare no competing interest.

\end{document}



\title{Bus bunching as a synchronisation phenomenon: Supplementary information}

\author{Vee-Liem Saw}
\affiliation{Division of Physics and Applied Physics, School of Physical and Mathematical Sciences, 21 Nanyang Link, Nanyang Technological University, Singapore 637371}
\affiliation{Data Science and Artificial Intelligence Research Centre, Block N4 \#02a-32, Nanyang Avenue, Nanyang Technological University, Singapore 639798}
\author{Ning Ning Chung}
\affiliation{Complexity Institute, 61 Nanyang Drive, Nanyang Technological University, Singapore 637335}
\author{Wei Liang Quek}
\affiliation{Division of Physics and Applied Physics, School of Physical and Mathematical Sciences, 21 Nanyang Link, Nanyang Technological University, Singapore 637371}
\author{Yi En Ian Pang}
\affiliation{Division of Physics and Applied Physics, School of Physical and Mathematical Sciences, 21 Nanyang Link, Nanyang Technological University, Singapore 637371}
\author{Lock Yue Chew}
\email{lockyue@ntu.edu.sg}
\affiliation{Division of Physics and Applied Physics, School of Physical and Mathematical Sciences, 21 Nanyang Link, Nanyang Technological University, Singapore 637371}
\affiliation{Data Science and Artificial Intelligence Research Centre, Block N4 \#02a-32, Nanyang Avenue, Nanyang Technological University, Singapore 639798}
\affiliation{Complexity Institute, 61 Nanyang Drive, Nanyang Technological University, Singapore 637335}
%

\date{\today}

\maketitle


\section{Analytical derivation of the phase transition to complete phase locking of all \texorpdfstring{$N$}{N} buses serving \texorpdfstring{$M$}{M} staggered bus stops in a loop}

Consider $N=2$ buses with natural angular frequencies $\omega_1>\omega_2$ ($\omega_i=2\pi f_i=2\pi/T_i$) serving $M=1$ bus stop in a loop. Suppose that the coupling $k:=s/l$ is strong enough such that these two buses are phase locked. (Recall that $s$ and $l$ are the people arrival and loading rates, respectively.) In that case, these two buses would always bunch at the bus stop and share the loading of people. Once everybody has been picked up, the two buses leave together, with the faster one pulling away. After one revolution, the faster one returns to the bus stop and begins picking up people. But before finishing, the slower one arrives (because $k$ is strong enough such that there are many people waiting at the bus stop) and the two buses share loading. These two buses are in such an equilibrium which repeats over and over.

\begin{figure}
\centering
\includegraphics[width=16cm]{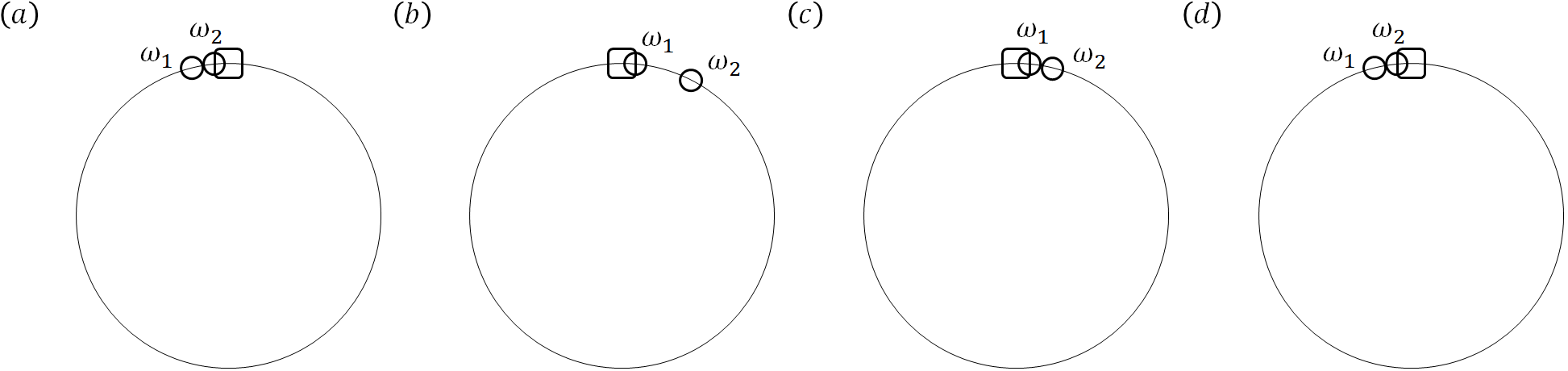}
\caption{$N=2$ buses serving $M=1$ bus stop in a loop, where $k$ is strong enough such that they are phase locked.}
\label{figS1}
\end{figure}

In Fig.\ \ref{figS1}, (a) is the moment when the two buses just leave the bus stop after picking up everybody, (b) is when the fast bus just arrives after one revolution, (c) is when the slow bus just arrives, and (d) is when both buses have finished picking up everybody and leave. The time elapsed from (a) to (b) is $T_1$, from (b) to (c) is $T_2-T_1$, from (c) to (d) is $\tau_\textrm{shared}$, where $\tau_\textrm{shared}$ is the duration when these two buses share loading. The total number of people to be picked up is $s$ times the total time elapsed from (a) to (d), which is $T_2+\tau_\textrm{shared}$. These people are picked up by:
\begin{enumerate}
\item Only the fast bus $=l(T_2-T_1)$.
\item Shared by the fast and slow buses $=2l\tau_\textrm{shared}$.
\end{enumerate}
The critical transition between no phase locking and phase locking is when $\tau_\textrm{shared}=0$. In that case,
\begin{align}
sT_2&=l(T_2-T_1)\\
k_c:&=\frac{s}{l}\\
&=1-\frac{T_1}{T_2}\\
&=1-\frac{f_2}{f_1}\\
&=1-\frac{\omega_2}{\omega_1}.
\end{align}

If there are $M$ staggered bus stops, then these sequence of events repeat at every bus stop. At the critical transition, $\tau_\textrm{shared}=0$ and the total number of people to pick up is still $sT_2$. However, the time interval for the fast bus to pick up all people is only $(T_2-T_1)/M$ since there are $M$ staggered bus stops in one revolution. Then, we have
\begin{align}
sT_2&=\frac{l(T_2-T_1)}{M}\\
k_c&=\frac{1}{M}\left(1-\frac{T_1}{T_2}\right)\\
&=\frac{1}{M}\left(1-\frac{f_2}{f_1}\right)\\
&=\frac{1}{M}\left(1-\frac{\omega_2}{\omega_1}\right).
\end{align}
So with $M$ bus stops, each bus stop multiplies the coupling strength. Hence, only one $M$-th of the coupling strength with one bus stop is required when there are $M$ bus stops.

Let us now consider $N$ buses with angular frequencies $\omega_1>\cdots>\omega_N$ serving $M=1$ bus stop in a loop, and we know that having $M$ staggered bus stops would be one $M$-th of $k_c$ for $M=1$. The total number of people to pick up is $s(T_N+\tau_\textrm{shared})$, since all buses have to wait for the slowest bus to reach the bus stop, and then all buses would share the load over the duration $\tau_\textrm{shared}$. These people are picked up by:
\begin{enumerate}
\item Only the first bus $=l(T_2-T_1)$.
\item Shared by only the first and second buses $=2l(T_3-T_2)$.
\item Shared by only the first, second and third buses $=3l(T_4-T_3)$.
\item $\cdots$
\item Shared by only the first $N-1$ buses $=(N-1)l(T_N-T_{N-1})$.
\item Shared by all buses $=Nl\tau_\textrm{shared}$.
\end{enumerate}
The critical transition between complete and partial phase locking is when $\tau_\textrm{shared}=0$. In that case,
\begin{align}
sT_N&=l\left[T_2-T_1+2T_3-2T_2+3T_4-3T_3+\cdots+(N-1)T_N-(N-1)T_{N -1}\right]\\
&=l\left[(N-1)T_N-\sum_{i=1}^{N-1}{T_i}\right]\\
k_c&=\sum_{i=1}^{N-1}{\left(1-\frac{T_i}{T_N}\right)}.
\end{align}
Thus, the critical transition between complete and partial phase locking for the general case of $N$ buses serving $M$ staggered bus stops in a loop is:
\begin{align}
k_c&=\frac{1}{M}\sum_{i=1}^{N-1}{\left(1-\frac{\omega_N}{\omega_i}\right)},
\end{align}
where $\omega_N/\omega_i=f_N/f_i=T_i/T_N$.
